\documentstyle[multicol,aps,psfig]{revtex}
\begin{document} 
\draft
\title{Equation of motion approach for the Hubbard model:\\
improved decoupling scheme, charge fluctuations,
and the metal-insulator transition}
\author{Avinash Singh}
\address{Department of Physics, Indian Institute of Technology, Kanpur-208016, India}
\maketitle
\begin{abstract} 
A new decoupling scheme is developed for the 
Hubbard model which provides a unified description
of the spin-symmetric (paramagnetic metallic and insulating) phases
as well as the broken-symmetry AFI phase. 
Independent of magnetic ordering, the scheme yields,
in the lowest order, 
the correct strong-coupling bandwidth of order $J$
(for the NN hopping model)
and band gap of order $U$,
a non-zero critical interaction strength (above which the band gap opens)
only if same-sublattice hopping (e.g., NNN hopping) is also present,
and the correct integrated spectral weights (1/2 per spin)
in each band for the half-filled model.
The effects of charge and spin fluctuations,
including spin twisting due to finite spin correlation length,
are investigated within a static, random approximation.
A self-consistent evaluation of the disorder-averaged
self energy within the CPA is carried out numerically.
Fluctuations activate the hopping term at first order
resulting in band-broadening,
and the consequent decrease in the
band gap with fluctuation strength is obtained for
several $U$ values in two and three dimensions.
We find that the band gap shrinks to zero continuously, 
and subsequently the density of states $N(0)$ between the bands
grows continuously,
leading to a continuous metal-insulator transition.
For finite doping there is transfer of spectral weight between the
Hubbard bands, and a qualitative change in the nature of the
quasiparticle band dispersion, with an effective doping-induced
hopping strength and bandwidth of order $xt$.

\end{abstract}

\begin{multicols}{2}\narrowtext
\section{Introduction}
Electron correlation effects in narrow d- or f-bands\cite{fulde}
are manifested in a variety of remarkable macroscopic phenomena,
such as metal-insulator transition in transition-metal
compounds,\cite{mibook,mibook2}
magnetism and high-temperature superconductivity in cuprates,\cite{cuprates}
colossal magnetoresistance in manganates,
and non Fermi-liquid behavior in transition-metal oxides,\cite{nfl_tmo}
high-T$_{\rm c}$ superconductors,\cite{nfl_hightc}
and heavy-fermion metals.\cite{nfl_hfm}
Incorporating electronic correlation in essence,
the prototypical Hubbard model,
theoretically studied first in some detail by
Hubbard,\cite{hubbard_1,hubbard_2,hubbard_3}
has been therefore intensively studied in recent years in these contexts.
While much attention has also been devoted to the doped region
away from half-filling,\cite{cuprates} motivated by the discovery of 
high-T$_{\rm c}$ superconductivity in the doped cuprates,\cite{hightc}
the discussion in the following is mainly limited to the half-filled
case, focusing only on the magnetic and metal-insulator
aspects of the phase diagram. 

Even at half-filling the Hubbard model in three dimensions has
a fairly rich phase diagram in the interaction-temperature
($U-T$) space, consisting of antiferromagnetic insulating (AFI),
paramagnetic insulating (PI), and paramagnetic metallic (PM)
phases.\cite{mibook} In addition, in the absence of Fermi-surface nesting
(for instance due to next-nearest-neighbor (NNN)
hopping on a bipartite lattice)
a gapless antiferromagnetic metallic (AFM) phase
also appears possible.\cite{gapless} 
The widely different nature of the constituent phases
(metallic and insulating, magnetic and nonmagnetic) 
has made it exceedingly difficult for the entire phase diagram
to be studied within a single theoretical framework, and over the years
various approaches have been developed to study different parts.
The brief account given below of some of the recent developments
will also serve to highlight the different physical processes
of importance in different sections of the phase diagram.

The study of magnetism in correlated electron systems goes back 
quite some time, to the theory of spin-waves in the band model
of ferromagnetic metals.\cite{band} The correlation effect is implicit
in the spin-wave excitation, which can be regarded as a bound state
of an electron of one spin with a hole of opposite spin.
Propelled by the discovery of nearly two-dimensional
antiferromagnetism in ${\rm La_2CuO_4}$,\cite{vaknin}
the intense activity in recent years
in low-dimensional magnetism and spin-fluctuation
effects, and more recently in three-dimensional magnetism,
has led to significant progress.

Starting with the AFI phase,
a major element in this progress has been the unified
understanding of this phase in the whole $U/t$ range
within a many-body-theoretical framework,
thus providing a bridge between the weak-coupling
Slater spin-density-wave limit and the strong-coupling
Heisenberg limit of localized spins.
For a generalized Hubbard model with multiple
degenerate orbitals per site,
it has been shown diagrammatically that 
within a systematic inverse-degeneracy expansion,
which preserves the spin-rotational symmetry
and hence the Goldstone mode order by order,
identical results are obtained
in the strong coupling limit ($U/t \rightarrow\infty$)
for all quantum corrections, order by order,
as from the linear spin-wave analysis of the equivalent
quantum Heisenberg antiferromagnet (QHAF).\cite{quantum}
Various quantities characterizing the AFI phase
have been evaluated in the whole $U/t$ range,
both in two and three dimensions,
and shown to interpolate properly
between the weak and strong coupling limits.
These include the spin-wave (magnon) velocity,\cite{swz,singh.tesanovic}
magnon energy and density of states,\cite{spectral,logan_dos}
sublattice magnetization $m$,\cite{swz,logan_dos,glswa,trans}
and the local transverse spin fluctuation magnitude 
$\langle S_\perp ^2 \rangle \equiv
\langle S_i ^- S_i ^+ + S_i ^+ S_i ^- \rangle$.\cite{trans,sp_exc}
The rapid suppression (with increasing $U$)
in the contribution of
particle-hole excitations across the charge gap,
and the rapid rise in that of the low-energy collective (magnon) excitations
indicates that an effective spin picture is approximately
valid down to surprisingly low $U$ values.\cite{sp_exc}

For the three-dimensional antiferromagnet,
the $U$-dependence of the N\'{e}el temperature $T_{\rm N}$,
which determines the magnetic phase boundary between the ordered
AFI phase and the spin disordered PI phase,
has also been quantitatively studied in the whole $U/t$ range.
A variety of techniques have been employed including
functional integral formalism, mainly within the static 
approximation,\cite{kakehashi,hasegawa,moriya,moriya2}
quantum Monte Carlo methods,\cite{hirsch,scalettar}
dynamical mean-field theory,\cite{dmft_neel,dmft_review}
which becomes exact in the limit of infinite dimensions,\cite{vollhardt} 
Onsager reaction field theory,\cite{orf}
self-consistent spin-fluctuation theory,\cite{neel} etc. 
Except for the last two,
all share the common feature of yielding a $T_{\rm N}$
which approaches, in the strong coupling limit,
the mean-field-theory (MFT) result 
for the equivalent spin-1/2 QHAF. 
This is due to the neglect of the long-wavelength,
low-energy magnon excitations,
which reduce $T_{\rm N}$ to nearly 70 \% of the MFT value
within the spin-fluctuation theory,\cite{neel}
in excellent agreement with recent QMC results for the
QHAF.\cite{sandvik}

The $U$-dependence of $T_{\rm N}$
has been shown to be very closely related to that
of the maximum magnon energy $\omega_{\rm m}$
which sets the magnon energy scale.\cite{neel}
Starting from the strong coupling limit
and going down in interaction strength,  $\omega_{\rm m}$
initially increases as $J=4t^2/U$, the exchange coupling.
However, as the weak coupling limit is approached
the falling charge gap ($2\Delta$) compresses the magnon spectrum,
and when $2\Delta \sim \omega_{\rm m}$ the energy scale
$\omega_{\rm m}$ turns over and starts falling,
eventually approaching $2\Delta$ in the weak coupling limit.

We first consider traversing up in temperature,
starting in the ordered AFI phase in the intermediate coupling regime.
With increasing temperature thermal excitation of the low-energy magnons
enhances the transverse spin fluctuation
$\langle S_\perp ^2 \rangle$, thereby reducing the AF order parameter,
until at $T=T_N \sim \omega_{\rm m}$, the sublattice magnetization
$\langle S_i ^z \rangle _A $ vanishes and the staggered
susceptibility diverges, marking the onset of the PI phase.
With further increase in temperature beyond $T_N$,
since no further reduction in the sublattice magnetization is possible,
a low-momentum cutoff must be introduced in the
sum over magnon modes.\cite{neel} The corresponding length scale
sets an upper limit to the magnon wavelength, and therefore
determines the  spin correlation length $\xi$ such that
AF order exists only upto length scale $\xi$.
The PI phase is thus characterized by an energy gap,
a spin correlation length, and local magnetization
$\langle \vec{S_i} \rangle$ whose direction slowly fluctuates
in space and time.

The twisting between neighboring spins relative to the
AF alignment increases as $\xi ^{-1}$ with temperature,
and this twist activates the hopping term at first order,
resulting in a broadening of the two bands,
as in the spiral phase studied earlier in the context of the
doped Hubbard model.\cite{spiral1,spiral2,spiral3}
The electronic bandwidth in the low-temperature ordered phase,
which is of the order of the exchange energy $J=4t^2/U$ due to
the second-order virtual hopping process,
is enhanced due to this twist by order $t\xi ^{-1}$,
thus providing a mechanism for coupling between the magnetic
(dis)ordering and the electronic spectrum. 
For $T >> J$, the correlation length $\xi \approx 0$,
and the bands are broadened to their maximum extent of order $W/2$,
the free electron half-bandwidth.
Finally when $T \approx U$, there is a crossover
into a semiconducting phase.

We now consider another sequence traversing down in interaction
strength, keeping the temperature fixed.
In this case charge fluctuations, which are strongly suppressed
in the strong coupling limit, progressively become more
important, and when $U \sim T$ the probability of a site being
doubly occupied or vacant becomes appreciable, and the
local moment defined by $m_i ^2 = \langle ( n_{i\uparrow} -
n_{i\downarrow} )^2 \rangle $ becomes insignificant.
Double occupancy and vacany also activate the hopping term
at first order, broadening the two bands.
As the band gap shrinks with decreasing $U$ and the bands broaden
due to this charge-fluctuation induced hopping,
the two bands eventually meet, resulting in the metal-insulator transtion.
Thus the disappearance of the local moment  and the 
transition into the metallic phase go hand-in-hand,
both driven by enhanced charge fluctuations.

An explicit demonstration of the existence of an 
insulating state with two separated bands,
{\em independent of any magnetic ordering},
which can undergo a transition with decreasing interaction strength
into a metallic phase was first provided by
Hubbard.\cite{hubbard_1,hubbard_3}
However, the decoupling scheme employed
within the equation of motion approach
resulted in several drawbacks
(discussed in more detail in section III), such as
inappropriate bandwidth, violation of the Fermi-liquid theory,
and a vanishing critical $U_c$ for the metal-insulator
phase transition for arbitrary hopping.
Of course, magnetic ordering and correlations,
and the associated low-energy magnetic excitations
were totally neglected. 

Another program,
which included the possibility of magnetic ordering,
was based on a static random one-body approximation to the
Hubbard model.\cite{economou}
The interaction term was replaced by an
effective, self-consistent random potential
with statistical correlation to simulate many-body
correlation effects. 
Only Ising-like magnetic excitations were included
and the low-energy magnon excitations were neglected. 
A continuous magnetic phase transition was obtained
between the AFI and the PI phases, whereas two crossover
temperatures were obtained around which, respectively, the DOS at
$E_{\rm F}$ becomes appreciable and the local moments
essentially disappear. 

In recent years the dynamical mean field theory,
which is exact in the limit of infinite dimensions,
has emerged as a controlled approximation for studying
the phase diagram in the whole parameter space.
In the limit of infinite dimensions
the interaction self energy becomes local,
and the quantum many-body problem
reduces to the self-consistent solution
of a single-site quantum problem
of an impurity embedded in an effective medium.
By extending this approach to two sites to account
for the sublattice structure in the AF state, 
a continuous magnetic phase transition was obtained,
as already mentioned above.
In addition, when $T_{\rm N}$ was sufficiently suppressed
(made possible by including a frustration-inducing NNN hopping term),
a first-order metal-insulator transition line,
ending in a second-order critical point was also
obtained.\cite{dmft_review}
Several associated features of the metal-insulator
transition in $\rm V_2 O_3$ have recently been discussed
within the DMFT approach, such as
volume discontinuities and resistivity changes,\cite{majumdar}
and optical conductivity on both sides of the transition.\cite{thomas}
The nature of the transition itself continues to be of interest,
and a recent study of the
changes in the density of states near the transition
indicates that the Fermi liquid breaks down before
the gap opens.\cite{schlipf}

While the AFI phase is well understood,
and spin fluctuation effects can be studied about the
broken-symmetry HF state even in the strong coupling limit,
for the magnetically-disordered PI and PM phases
a good analytical starting point at the same level
is still not available.
The decoupling approach presented here provides this
for studying the interplay of electronic spectral
properties and AF correlations in these states.
Furthermore, the approach is equally applicable to the
broken-symmetry state, allowing for a unified description
independent of magnetic ordering.

The paper is organized as follows.
After introducing the equation of motion approach
for the single-particle Green's function in section II,
and briefly reviewing the Hubbard I approximation in Section III,
improved decoupling schemes are presented in section IV
which take into account AF correlations.
Especially appropriate for the spin-disordered PI phase,
the final decoupling procedure in subsection C
offers several improvements over the Hubbard I approximation.
Fluctuations are included in section V
within a static random approximation
and a self-consistent evaluationt of the disorder-averaged
self energy within the CPA is carried out.
Fluctuation effects on the charge gap and the density of states
are examined in section VI
by studying the coupled, non-linear equations for the
self-consistently determined complex self energy.
Contact with known results for the broken-symmetry state is made
in section VII, showing the applicability of the decoupling scheme
irrespective of magnetic ordering.
Deviation from perfect AF ordering
and effects of short-ranged AF spin correlations on the electronic spectrum
are studied in section VIII.
The effects of hole doping on the qualitative nature of the
quasiparticle band dispersion,
and the transfer of spectral weight between the Hubbard bands
are briefly studied in section IX.
Some conclusions are presented in section X.

\section{Equation of motion approach}
We consider the Hamiltonian
\begin{equation}
H = H_0 + H_I = - \sum_{i\delta\sigma} 
t_\delta \; a_{i \sigma}^{\dagger} a_{i+\delta, \sigma}
+  U\sum_{i} n_{i \uparrow} \; n_{i \downarrow} \; ,
\end{equation}
where the kinetic energy term $H_0$ involves the hopping
matrix elements $t_\delta$ 
between sites $i$ and $i+\delta$, 
and $H_I$ is the local interaction term. 
In momentum space $H_0=\sum_{{\bf k}\sigma}
\epsilon_{\bf k} \;
a^\dagger _{{\bf k}\sigma}a_{{\bf k}\sigma}$,
where $\epsilon_{\bf k}= 
-\sum_{\delta} t_\delta \; e^{i{\bf k}.{\bf r}_\delta}$
denotes the lattice free-particle energy. 

We consider the equation of motion approach,
in which the time derivative of an operator
$\hat{O}(t)=e^{iHt}\hat{O}e^{-iHt}$ in the Heisenberg representation
is obtained from the commutator with the Hamiltonian, 
$i\partial_t \hat{O}(t)=[\hat{O}(t),H]= e^{iHt}[\hat{O},H]e^{-iHt}$,
and apply it to the 
time-ordered one-particle Green's function
\begin{eqnarray}
& & G_{ij}(t)=-i \langle \Psi | {\rm T} a_{i\uparrow}(t)
a_{j\uparrow}^\dagger (0) |\Psi\rangle = \nonumber \\
& & -i[\theta(t)\langle \Psi | a_{i\uparrow}(t)
a_{j\uparrow}^\dagger (0) |\Psi\rangle -
\theta(-t)\langle \Psi | a_{j\uparrow}(0)
a_{i\uparrow}^\dagger (t) |\Psi\rangle ] \; .
\end{eqnarray}
Taking the time derivative and using $\partial_t \theta(t)=
\delta(t)$, the anticommutation property $\{a_{i\uparrow},a_{j\uparrow}^\dagger\}=\delta_{ij}$,
and the commutator 
\begin{equation}
[a_{i\uparrow},H] = - \sum_\delta 
t_\delta \; a_{i+\delta\uparrow} 
+ U a_{i\uparrow}n_{i\downarrow} 
\end{equation} 
to determine $\partial_t a_{i\uparrow}(t)$, one obtains
\begin{eqnarray}
i\partial_t G_{ij}(t) &=& \delta(t)\delta_{ij} - 
\sum_\delta t_\delta \; G_{i+\delta,j}(t) \nonumber \\
&-& iU\langle \Psi | {\rm T} n_{i\downarrow}(t) a_{i\uparrow}(t)
a_{j\uparrow}^\dagger (0) |\Psi\rangle ,
\end{eqnarray}
which is an equation for $G$ in terms of a higher-order
correlation function.

Decoupling the equation of motion at this stage by
replacing the operator $n_{i\downarrow}$ in the last term
by its expectation value results in the Hartree-Fock
approximation
\begin{equation}
i\partial_t G_{ij}(t)= \delta(t)\delta_{ij} - 
\sum_\delta t_\delta \; G_{i+\delta,j}(t)
+U \langle n_{i\downarrow}\rangle G_{ij}(t)\; ,
\end{equation}
where the densities $\langle n_{i\sigma}\rangle$
are determined self consistently.
This approximation is a reasonable {\em starting point} 
in the broken-symmetry state,
when the long-time average of the local density $n_{i\downarrow}$
does not differ substantially from the short-time average.
For the AF state,
basic features such as the correlation effect and the associated
electron localization in the strong interaction limit
are actually already contained at this level, as evidenced
by the charge gap of order $U$ and the narrow AF
bandwidth of order $J$. 
Within a spin-fluctuation approach,
wherein transverse fluctuations about the HF state are  
systematically incorporated,
the AF state of the half-filled Hubbard model has been
extensively studied in recent years,
including the effects of disorder,
impurities, and vacancies.\cite{neel,dis,trans}
Even the strong coupling limit ($U/t \rightarrow \infty$)
is accessible within this formally weak-coupling expansion.
This has been explicitly demonstrated
by systematically evaluating
quantum corrections to various quantities
such as the magnon energy, AF order parameter, perpendicular
spin susceptibility, ground-state energy,
spin correlation length.\cite{quantum}

However, in the spin disordered state
this approximation breaks down for strong interaction,
when the expectation value of the local density 
$\langle n_{i\downarrow}\rangle =n/2$ (the mean density)
differs substantially from the short-time average
(either nearly zero or nearly one). 
In this case the local magnetization strongly fluctuates 
between 1 and -1, so that the average local magnetization 
$\langle m_i \rangle =
\langle n_{i\uparrow} - n_{i\downarrow}\rangle$ vanishes,
but the local moment $\mu_i$ defined by  
$\mu_i ^2 \equiv \langle m_i ^2 \rangle$ remains appreciable.
This strong local moment arises from the correlation effect
of an electron locally excluding the opposite-spin electron 
due to the strong electronic repulsion.

This provides a clue that the decoupling procedure can be improved
by ensuring that the operators which are replaced by their
averages involve minimal fluctuations. In the following we describe
several such decoupling procedures, all of which 
are quite distinct from that employed by Hubbard.\cite{hubbard_1}
This approximation, commonly referred to as the Hubbard I, 
is first briefly reviewed here for comparison.

\section{Hubbard I}
In Hubbard I the next step was to 
continue with the equation of motion approach 
and examine the time derivative of the 
last term in Eq. (4)
involving the self-energy correction,
$\Gamma_{ij}(t)=
-i U \langle \Psi | {\rm T} n_{i\downarrow}(t) a_{i\uparrow}(t)
a_{j\uparrow}^\dagger (0) |\Psi\rangle $.
Determining the time derivatives in  
$\partial_t [n_{i\downarrow}(t) a_{i\uparrow}(t)]$
from Eq. (3) and the commutator 
\begin{equation}
[n_{i\downarrow},H] = -\sum_\delta t_\delta \;
(a_{i\downarrow}^\dagger a_{i+\delta\downarrow} -
a_{i-\delta\downarrow}^\dagger a_{i\downarrow}) \; ,
\end{equation}
and using the anticommutation property 
$\{n_{i\downarrow} a_{i\uparrow} , a_{j\uparrow}^\dagger \}
=\delta_{ij}n_{i\downarrow}$, one obtains
\begin{eqnarray}
& & i\partial_t \Gamma_{ij}(t) = 
U \delta(t) \langle \Psi 
|\{n_{i\downarrow} a_{i\uparrow},a_{j\uparrow}^\dagger \} |\Psi\rangle
\nonumber \\
&+& U \langle \Psi | {\rm T}
\partial_t \{n_{i\downarrow}(t) a_{i\uparrow}(t)+
n_{i\downarrow}(t) a_{i\uparrow}(t) \}
 a_{j\uparrow}^\dagger (0) |\Psi\rangle \nonumber \\
&=& \delta(t)\delta_{ij} 
U\langle n_{i\downarrow}\rangle 
+ iU \sum_\delta t_\delta \;\langle \Psi | {\rm T} 
n_{i\downarrow}(t)  a_{i+\delta\uparrow}(t)
a_{j\uparrow}^\dagger (0) |\Psi\rangle 
\nonumber \\
&-& i U^2 \langle \Psi |{\rm T}
n_{i\downarrow} ^2 (t) a_{i\uparrow}(t)
a_{j\uparrow}^\dagger (0) |\Psi\rangle 
+ iU \sum_\delta t_\delta \; \langle \Psi | {\rm T} \nonumber \\
& &\{a_{i\downarrow}^\dagger (t) a_{i+\delta\downarrow}(t)
- a_{i-\delta \downarrow}^\dagger (t)
a_{i\downarrow}(t) \} a_{i\uparrow}(t)
a_{j\uparrow}^\dagger (0) |\Psi\rangle  \; .
\end{eqnarray}

In order to obtain a closed 
equation for $\Gamma$, the equations were decoupled 
at this stage by noticing that 
$n_{i\downarrow} ^2 = n_{i\downarrow}$,
and replacing the operators $n_{i\downarrow}(t)$,
and $\{a_{i\downarrow}^\dagger (t) a_{i+\delta\downarrow}(t) 
- a_{i-\delta \downarrow}^\dagger (t) a_{i\downarrow}(t) \} $
in the last three terms by their expectation values
in a spin-symmetric, translationally-invariant ground state, with 
$\langle a_{i\downarrow}^\dagger (t) a_{i+\delta\downarrow}(t) 
- a_{i-\delta \downarrow}^\dagger (t) a_{i\downarrow}(t) 
\rangle = 0$, and
$\langle n_{i\downarrow}\rangle = \langle n_{i\uparrow}\rangle = 
n/2$, the average density for each spin.
Fourier transforming to frequency and momentum space
using
\begin{eqnarray}
G_{ij}(t) &=& \sum_{\bf k}
\int\frac{d\omega}{2\pi} G({\bf k}\omega)e^{i({\bf k}.
{\bf r}_{ij}-\omega t)} \nonumber \\
\delta_{ij} \delta(t)&=& \sum_{\bf k}\int\frac{d\omega}{2\pi} e^{i({\bf k}.{\bf r}_{ij}-\omega t)}
\nonumber \\
-\sum_\delta t_\delta \; \delta_{i+\delta,j} 
&=& \sum_{\bf k}  \epsilon_{\bf k} \; 
e^{i{\bf k}.{\bf r}_{ij}} 
\nonumber \\
-\sum_\delta t_\delta \; G_{i+\delta,j}(t) &=& 
\sum_{\bf k} \int\frac{d\omega}{2\pi} G({\bf k}\omega)
\epsilon_{\bf k}  \;e^{i({\bf k}.{\bf r}_{ij}-\omega t)} \; ,
\end{eqnarray}
one obtains
\begin{equation}
\omega\Gamma({\bf k}\omega)=\frac{Un}{2} 
[ 1+ \epsilon_{\bf k} G({\bf k}\omega) ] 
+ U \Gamma({\bf k}\omega) 
\; .
\end{equation}
Substituting for $\Gamma({\bf k}\omega)=\frac{Un}{2}
[1+\epsilon_{\bf k} G({\bf k}\omega)]/
(\omega -U)$ from above into Eq. (4), 
and solving for $G({\bf k}\omega)$
one obtains the Hubbard I result
\begin{equation}
G({\bf k}\omega)=\frac{\omega-U(1-n/2)}
{(\omega-\epsilon_{\bf k})(\omega-U) - Un\epsilon_{\bf k} /2 }\; .
\end{equation}
In the half-filled case ($n=1)$ this can be written as, 
\begin{equation}
G({\bf k}\omega)= \frac{1}{2} \left [
\frac{1-\frac{\epsilon_{\bf k}}{E_{\bf k}}}
{\omega -E_{\bf k} ^{(1)} } +
\frac{1+\frac{\epsilon_{\bf k}}{E_{\bf k}}}
{\omega -E_{\bf k} ^{(2)}} 
\right ] \; ,
\end{equation}
where $E_{\bf k}= \sqrt{U^2 + \epsilon_{\bf k} ^2}$, and
$E_{\bf k} ^{(1,2)}=[(\epsilon_{\bf k} +U)
\mp E_{\bf k} ]/2$ are the two band energies.
In the strong coupling limit
these band energies reduce to $\epsilon_{\bf k} /2$ and 
$U+\epsilon_{\bf k} /2$ to lowest order, 
giving a bandwidth of $W/2$,
where $W$ is the free-particle bandwidth.
Thus in going from the non-interacting limit to
the strong-coupling limit, the bandwidth decreases
only from $W$ to $W/2$, whereas one expects a 
correlation-narrowed bandwidth of order $J$ for the NN 
hopping model. This is a serious drawback
of the Hubbard I approximation in the strong correlation limit, 
and can be traced to the replacement of $n_{i\downarrow}$, 
a strongly fluctuating quantity, by its average value.
Another drawback pertains to the absence of a critical
interaction strength for the energy gap. 
For a general free-particle energy $\epsilon_{\bf k}$, 
the electronic spectrum is split into two bands 
for all $U$. Also, in general,
the integrated spectral weight in each Hubbard band
is not exactly one-half per spin, although the total for both
bands is identically one. 
This amounts to an inconsistency in the insulating state
in that the number of states in $\bf k$-space 
lying below the Fermi energy does not 
exactly yield the number of particles.

\section{Improved decoupling schemes}
We adopt a different approach here 
which removes all of the drawbacks cited above. 
We continue with Eq. (4) for $\partial_t G_{ij}$
and consider the second derivative
\begin{eqnarray}
i\partial_t ^2 G_{ij}(t) &=& 
\partial_t \delta(t)\delta_{ij} - \sum_\delta t_\delta \; 
\partial_t G_{i+\delta,j}(t) \nonumber \\
&-& iU\; \langle \Psi | {\rm T}
\partial_t [n_{i\downarrow}(t) a_{i\uparrow}(t)]
a_{j\uparrow}^\dagger (0) |\Psi\rangle \nonumber \\
 &-& iU \delta(t)\; \langle \Psi | \{n_{i\downarrow} a_{i\uparrow} , 
a_{j\uparrow}^\dagger \} |\Psi\rangle .
\end{eqnarray}
The derivative of $G_{i+\delta,j}(t)$ is substituted from 
Eq. (4), the derivative 
$\partial_t [n_{i\downarrow}(t) a_{i\uparrow}(t)]$
obtained as before from Eqs. (3) and (6),
and in the last term  
$\{n_{i\downarrow} a_{i\uparrow} , a_{j\uparrow}^\dagger \}
=\delta_{ij}n_{i\downarrow}$.
Putting all this together, and multiplying by $i$,
we obtain
\begin{eqnarray}
& & i^2 \partial_t ^2 G_{ij}(t)=
i \partial_t \delta(t)\delta_{ij} 
+ \delta(t) [\delta_{ij}U\langle n_{i\downarrow}\rangle \nonumber \\
&-& \sum_\delta t_\delta \; \delta_{i+\delta,j} ] 
+ \sum_{\delta \delta'} t_\delta \; t_{\delta '} \; 
G_{i+\delta+\delta' ,j} \nonumber \\
&+& iU \sum_\delta t_\delta \; \langle \Psi | {\rm T} 
\{n_{i\downarrow}(t)+ n_{i+\delta \downarrow}(t) \}
a_{i+\delta\uparrow}(t)
a_{j\uparrow}^\dagger (0) |\Psi\rangle \nonumber \\
 &-& i U^2 \;\langle \Psi |{\rm T} 
n_{i\downarrow} ^2 (t) a_{i\uparrow}(t)
a_{j\uparrow}^\dagger (0) |\Psi\rangle 
+ iU \sum_\delta t_\delta \; \langle \Psi | {\rm T} \nonumber \\
& & \{a_{i\downarrow}^\dagger (t) 
a_{i+\delta\downarrow}(t) 
- a_{i-\delta \downarrow}^\dagger (t) 
a_{i\downarrow}(t) \} a_{i\uparrow}(t)
a_{j\uparrow}^\dagger (0) |\Psi\rangle \; .
\end{eqnarray}

We now proceed further along three different routes
by introducing three decoupling schemes termed A, B, and C
in the following subsections.
Schemes A and B are most suitable  
for hopping terms which only connect sites of opposite 
and same sublattices respectively. 
The best features of A and B are 
then combined in the final scheme C which is applicable
for any general hopping. However, as we shall see, we are
constrained to the consideration of a bipartite lattice.  
\subsection{}
Adding $- nU\; 
i\partial_t G_{ij}(t) + (nU/2)^2 G_{ij}(t)$
to both sides of Eq. (13), 
where $n=\langle n_{i\uparrow}+n_{i\downarrow}\rangle $ is
the total density, and substituting for
$\partial_t G_{ij}(t)$ from Eq. (4), we obtain
\begin{eqnarray}
& & i^2 \partial_t ^2 G_{ij}(t)
- nU\; i\partial_t G_{ij}(t) +(nU/2)^2 G_{ij}(t)
= i\partial_t \delta(t)\delta_{ij} \nonumber \\
&+& \delta(t) [\delta_{ij}
U\{\langle n_{i\downarrow}\rangle -n\}
- \sum_\delta t_\delta \; \delta_{i+\delta,j} ] 
+ \sum_{\delta \delta '} t_\delta \; 
t_{\delta '} G_{i+\delta+\delta' ,j} \nonumber \\
 &+& iU\sum_\delta t_\delta \; \langle \Psi | {\rm T} 
\{n_{i\downarrow}(t)+ n_{i+\delta \downarrow}(t) - n \}
 a_{i+\delta\uparrow}(t)
a_{j\uparrow}^\dagger (0) |\Psi\rangle \nonumber \\
 &-& iU^2 \; \langle \Psi |{\rm T} 
\{n_{i\downarrow}(t)-n/2 \} ^2 a_{i\uparrow}(t)
a_{j\uparrow}^\dagger (0) |\Psi\rangle 
+ iU\sum_\delta t_\delta \;  \nonumber \\
& & \langle \Psi | {\rm T}  
\{a_{i\downarrow}^\dagger (t) 
a_{i+\delta\downarrow}(t) 
- a_{i-\delta \downarrow}^\dagger (t) 
a_{i\downarrow}(t) \} a_{i\uparrow}(t)
a_{j\uparrow}^\dagger (0) |\Psi\rangle  \; . \nonumber \\
\end{eqnarray}

In order to decouple this equation for $G$ involving several
higher-order correlation functions we now replace the operators 
$\{n_{i\downarrow}(t)+ n_{i+\delta \downarrow}(t) - n \}$,
$\{n_{i\downarrow}(t) -n/2\} ^2 $, and 
$\{a_{i\downarrow}^\dagger (t) a_{i+\delta\downarrow}(t) 
- a_{i-\delta \downarrow}^\dagger (t) a_{i\downarrow}(t) \} $
in the last three terms by their ground-state expectation values,
again in the spin-symmetric, translationally-invariant ground state, 
with $\langle n_{i\downarrow}\rangle =
\langle n_{i\uparrow}\rangle = n/2$, the average density for each spin.
At half-filling ($n=1$), we have 
$\{n_{i\downarrow}(t)-1/2 \} ^2 = (1/2)^2$, a c-number, so that
no approximation is involved in this replacement.
Furthermore, in the strong coupling limit, when double occupancy and 
vacancy  are both prohibited, the total charge density operator
$n_{i\uparrow}+n_{i\downarrow} = 1$, the unit operator,
so that the term
$\{n_{i\downarrow}(t)-1/2 \} ^2 = m_i ^2 /4$
is also related to the second moment of the local
magnetization $m_i$.
In general, $n_{i\downarrow}(t)-1/2$ fluctuates between the two 
values $m_i/2 = \mp 1/2$, depending on whether the site 
is occupied or empty.

The average 
$\langle n_{i\downarrow}+ n_{i+\delta \downarrow} - n \rangle$
vanishes in the paramagnetic state
when $\langle n_{i\downarrow}\rangle = \langle 
n_{i+\delta \downarrow} \rangle =n/2$. More importantly, it
also vanishes in the presence of AF correlations,
provided the sites $i$ and $i+\delta$ are in opposite sublattices,
in which case 
$\langle n_{i+\delta \downarrow} \rangle =
\langle n_{i \uparrow} \rangle$. 
This scheme is therefore most appropriate when the hopping term
$t_\delta$ connects sites $i$ and $i+\delta$ of opposite
sublattices, as in the NN hopping case.
We take
$\langle a_{i\downarrow}^\dagger (t) a_{i+\delta\downarrow}(t) 
- a_{i-\delta \downarrow}^\dagger (t) a_{i\downarrow}(t) 
\rangle = 0$, as in Hubbard I.
This is obviously so if translational symmetry is present,
but also in presence of AF correlations, as both terms are
off-diagonal in the sublattice basis, and of equal magnitude.
After Fourier transformation to frequency and momentum space,
and introducing a Hartree shift in the energy axis through the
transformation $\omega-nU/2 \rightarrow \omega$,
we finally obtain
\begin{equation}
G(k\omega) = \frac{\omega +\epsilon_k}
{\omega^2 - (\Delta^2 + \epsilon_k ^2)} 
= 
\frac{1}{2} \left [
\frac{1+\frac{\epsilon_k}{E_k}}{\omega  - E_k  } +
\frac{1-\frac{\epsilon_k}{E_k}}{\omega  + E_k  } 
\right ] \; ,
\end{equation}
where $E_k=\sqrt{\Delta^2 +\epsilon_k ^2}$ and
$\Delta^2 = U^2 \langle \{n_{i\downarrow}(t) - 1/2 \} ^2 
\rangle = U^2 /4$. 
The energy spectrum thus splits into two bands,
independently of whether there is long-range AF order or not,
with an energy gap of $2\Delta=U$.
Furthermore, the bandwidth is correctly of order $J$
in the strong coupling limit.
\subsection{} 
An alternate decoupling scheme
is more appropriate when the hopping term
involves sites in the same sublattice.
Adding to both sides of Eq. (13), the terms  
$-nU\; i \partial_t G_{ij} +(nU/2)^2 G_{ij}
-nU\sum_\delta t_\delta \; G_{i+\delta,j}
+2i\sum_\delta t_\delta \; \partial_t G_{i+\delta,j}
+\sum_{\delta \delta '} t_\delta \; t_{\delta '}\; 
G_{i+\delta+\delta ',j}$,
and substituting for $\partial_t G(t)$ from Eq. (4), we obtain
\begin{eqnarray}
& & 
i^2 \partial_t ^2 G_{ij} -nU \; 
i\partial_t G_{ij} +(nU/2)^2 G_{ij} \nonumber \\
&+& 2i\sum_\delta t_\delta \;\partial_t G_{i+\delta,j}
-nU\sum_\delta t_\delta \; G_{i+\delta,j}
+\sum_{\delta\delta '} t_\delta \; t_{\delta'}
G_{i+\delta+\delta ',j} \nonumber \\
&=& 
i\partial_t \delta(t)\delta_{ij} + \delta(t) [\delta_{ij}
U\{\langle n_{i\downarrow}\rangle -n\}
+\sum_\delta t_\delta \; \delta_{i+\delta,j} ] \nonumber \\
 &+& iU\sum_\delta t_\delta \; \langle \Psi | {\rm T} 
\{n_{i\downarrow}(t)- n_{i+\delta \downarrow}(t) \}
 a_{i+\delta\uparrow}(t)
a_{j\uparrow}^\dagger (0) |\Psi\rangle \nonumber \\
 &-& iU^2\; \langle \Psi |{\rm T} 
\{n_{i\downarrow}(t) -n/2 \} ^2  a_{i\uparrow}(t)
a_{j\uparrow}^\dagger (0) |\Psi\rangle 
+ iU\sum_\delta t_\delta \; \nonumber \\
& & \langle \Psi | {\rm T} 
\{a_{i\downarrow}^\dagger (t) a_{i+\delta\downarrow}(t) 
- a_{i-\delta \downarrow}^\dagger (t) 
a_{i\downarrow}(t) \} a_{i\uparrow}(t)
a_{j\uparrow}^\dagger (0) |\Psi\rangle  \; .
\end{eqnarray}
We notice that on the right-hand side the second-order 
hopping term $(t_\delta \; t_{\delta '})$ present in 
Eq. (15) has exactly cancelled.
Again, as before, the operators are replaced by their expectation
values.
We note that we can safely set $\langle n_{i\downarrow}-
n_{i+\delta \downarrow}\rangle =0$,
not only in the symmetric paramagnetic state,
but even in the presence of AF correlations, provided 
that sites $i$ and $i+\delta$ are in the same sublattice.
Fourier transformation leads to 
\begin{equation}
(\omega -nU/2 -\epsilon_k)^2 G(k\omega) =
(\omega -nU/2 -\epsilon_k) + \Delta^2 G(k\omega) \; .
\end{equation}
With the same Hartree shift employed through the transformation
$\omega -nU/2 \rightarrow \omega$,  we obtain
\begin{eqnarray}
G(k\omega)&=& \frac{\omega -\epsilon_k}
{(\omega -\epsilon_k )^2  - \Delta^2 } \nonumber \\
&=& \frac{1}{2} \left [
\frac{1}{\omega -\epsilon_k -\Delta} +
\frac{1}{\omega -\epsilon_k +\Delta} \right ] \; .
\end{eqnarray}
Thus $\Delta$ causes the
free-particle spectrum to be split
into two bands, with no correlation effect on the bandwidth.
The two bands overlap when the band separation
$2\Delta=U$ is smaller than the free particle bandwidth,
but a gap opens up beyond a certain critical interaction 
strength. Hence for a general $\epsilon_{\bf k}$, there is
a non-zero critical interaction strength. However, the number
of states in each band is precisely 1/2 per spin.   
\subsection{}
Finally, we consider a third scheme which combines the
best features of A and B. 
We recall the replacement of operators by their expectation values
in the decoupling procedures, and focus on the operators
$\{n_{i\downarrow}(t) + n_{i+\delta \downarrow}(t) - n\}$ and 
$\{n_{i\downarrow}(t)- n_{i+\delta \downarrow}(t) \}$
in Eqs. (14) and (16), 
particularly in view of AF correlations in the ground state. 
For $\delta$ connecting sites of opposite sublattices,
the short-time average 
$\langle n_{i\downarrow}(t)+ n_{i+\delta \downarrow}(t) - n 
\rangle $ is weakly fluctuating, but not so when
sites $i$ and $i+\delta$ are in the same sublattice.
Therefore, scheme A is suitable when hopping
is limited to opposite sublattices.
On the other hand, the short-time average 
$\langle n_{i\downarrow}(t)- n_{i+\delta \downarrow}(t) 
\rangle $ is weakly fluctuating when $\delta$ connects sites
of the same sublattice,
so that scheme B is suitable when hopping is limited
to the same sublattice. 
This immediately suggests that for general hopping
both schemes can be 
further improved upon by taking an intermediate measure.  

For this purpose we are constrained to the 
consideration of a bipartite lattice, for which the 
hopping terms can be divided into two groups, 
connecting sites of opposite and same sublattices. 
We therefore write,
\begin{eqnarray}
H_0 &=& - \sum_{i\lambda\sigma} 
t_\lambda a_{i \sigma}^{\dagger} a_{i+\lambda, \sigma}
 - \sum_{i\mu\sigma} 
t_\mu a_{i \sigma}^{\dagger} a_{i+\mu, \sigma} \nonumber \\
&=& \sum_{\bf k \sigma} (\epsilon_{\bf k \lambda} 
+ \epsilon_{\bf k \mu}) a_{\bf k \sigma}^\dagger
a_{\bf k \sigma}
\end{eqnarray}
where sites $i+\lambda$ and $i+\mu$ refer to neighbors 
of site $i$ in the opposite and same sublattices respectively,
and furthermore
$\epsilon_{\bf k \lambda}=-\sum_\lambda t_\lambda 
e^{i{\bf k}.{\bf r}_\lambda }$
and $\epsilon_{\bf k \mu}=-\sum_\mu t_\mu 
e^{i{\bf k}.{\bf r}_\mu }$
are the two associated free-particle energies.

The intermediate measure which leads to 
an improvement over both schemes A and B is 
simply effected by including only those additional
terms of scheme B which correspond to the same
sublattice. Thus,   
we add to both sides of Eq. (13) the terms  
$ -nU\; i \partial_t G_{ij} + (nU/2)^2 G_{ij}
-nU\sum_\mu t_\mu \; G_{i+\mu,j}
+2\; i\sum_\mu t_\mu \; \partial_t G_{i+\mu,j}
+\sum_{\mu \mu '} t_\mu \; t_{\mu '}\; 
G_{i+\mu+\mu ',j}$,
where the sum over neighbors $(\mu,\mu')$ 
in the last three terms
are restricted to sites in the {\em same} sublattice.
As before, we obtain after substituting for 
$\partial_t G$ from Eq. (4),
\begin{eqnarray}
& & i^2\partial_t ^2 G_{ij}
-nU \; i\partial_t G_{ij}
+(nU/2)^2 G_{ij} + 2i\sum_\mu t_\mu \;\partial_t G_{i+\mu,j}
\nonumber \\
&-& nU\sum_\mu t_\mu \; G_{i+\mu,j}
+\sum_{\mu\mu '} t_\mu \; t_{\mu'}
G_{i+\mu+\mu ',j} \nonumber \\
&=& i\partial_t \delta(t)\delta_{ij}
+ \delta(t) [\delta_{ij}
U\{\langle n_{i\downarrow}\rangle -n\} \nonumber \\
&+& \sum_\mu t_\mu \; \delta_{i+\mu,j} 
-\sum_\lambda t_\lambda \; \delta_{i+\lambda,j}]
+\sum_{\lambda,\lambda'} t_\lambda \; t_{\lambda'}
G_{i+\lambda+\lambda',j}  \nonumber \\
 &+& iU\sum_\mu t_\mu \; \langle \Psi | {\rm T} 
\{n_{i\downarrow}(t)- n_{i+\mu \downarrow}(t) \}
 a_{i+\mu\uparrow}(t)
a_{j\uparrow}^\dagger (0) |\Psi\rangle \nonumber \\
 &+& iU\sum_\lambda t_\lambda \; \langle \Psi | {\rm T} 
\{n_{i\downarrow}(t) + n_{i+\lambda \downarrow}(t) -n \}
 a_{i+\lambda\uparrow}(t)
a_{j\uparrow}^\dagger (0) |\Psi\rangle \nonumber \\
 &-& iU^2\; \langle \Psi |{\rm T} 
\{n_{i\downarrow}(t) -n/2 \} ^2  a_{i\uparrow}(t)
a_{j\uparrow}^\dagger (0) |\Psi\rangle 
+ iU\sum_\delta t_\delta \; \nonumber \\
& & \langle \Psi | {\rm T} 
\{a_{i\downarrow}^\dagger (t) a_{i+\delta\downarrow}(t) 
- a_{i-\delta \downarrow}^\dagger (t) 
a_{i\downarrow}(t) \} a_{i\uparrow}(t)
a_{j\uparrow}^\dagger (0) |\Psi\rangle  \; .
\end{eqnarray}
The cancellation of the second-order hopping terms,
referred to earlier below Eq. (16),
is only partial now, and only terms involving opposite
sublattice hopping finally survive. 

Again replacing the operators with their expectation
values as before, with 
$\langle n_{i\downarrow}- n_{i+\mu \downarrow}\rangle =
\langle n_{i\downarrow}+ n_{i+\lambda \downarrow} 
- n \rangle =0$,
we obtain after Fourier transformation 
\begin{eqnarray}
(\omega -nU/2 -\epsilon_{\bf k \mu})^2 G({\bf k}\omega)
&=&
\omega -nU/2 -\epsilon_{\bf k \mu} + 
\epsilon_{\bf k \lambda} \nonumber \\
&+& (\Delta^2 + \epsilon_{\bf k \lambda} ^2) G({\bf k}\omega)
\end{eqnarray}
which, after the same Hartree shift 
$\omega -nU/2 \rightarrow \omega$, yields
\begin{eqnarray}
G({\bf k}\omega)&=& \frac{(\omega -\epsilon_{\bf k \mu}) 
+ \epsilon_{\bf k \lambda}}
{(\omega -\epsilon_{\bf k \mu} )^2  - 
E_{\bf k \lambda} ^2 } \nonumber \\
&=& \frac{1}{2} \left [
\frac{1+\frac{\epsilon_{\bf k \lambda}}
{E_{\bf k \lambda}}}{\omega -\epsilon_{\bf k \mu} -
E_{\bf k \lambda}} +
\frac{1-\frac{\epsilon_{\bf k \lambda}}
{E_{\bf k \lambda}}}{\omega -\epsilon_{\bf k \mu} +
E_{\bf k \lambda}}  \right ]
\end{eqnarray}
where $E_{\bf k \lambda} = \sqrt{\Delta^2 + 
\epsilon_{\bf k \lambda} ^2}$
is associated with the opposite sublattice hopping.
Results of both schemes A and B are recovered 
by taking appropriate limits. 
For only opposite sublattice hopping 
$\epsilon_{\bf k \mu} = 0$, so that Eq. (15) is obtained,
whereas for only same sublatttice hopping 
$\epsilon_{\bf k \lambda} = 0$, which yields Eq. (18). 
In general, when hopping involving same sublattice sites
(e.g. NNN hopping) is present,
a finite $\Delta$ is required for the band gap
to appear, yielding a critical interaction strength
which separates the metallic and insulating phases.

Quite generally, $\epsilon_{\bf k \lambda}$,
the free-particle band energy associated with opposite-sublattice
hopping, changes sign under the transformation
${\bf k}\rightarrow {\bf k} + \mbox{\boldmath $\pi$}$, 
while the band energy $\epsilon_{\bf k \mu} \mp
E_{\bf k \lambda}$ does not. Therefore,
the ${\bf k}$-space is split into two degenerate zones.
For evaluation of the integrated spectral weight,
the ${\bf k}$-sum therefore covers the whole ${\bf k}$ space.
As $\epsilon_{\bf k \lambda}$ is odd, 
the integrated spectral weight in each band
is identically one-half per spin.

\section{Fluctuation effects}
In the previous section, the equation of motion for $G_{ij}(t)$
was decoupled by replacing the operators
$\{n_{i\downarrow}+n_{i+\lambda\downarrow} -n \}$,
$\{n_{i\downarrow} - n_{i+\mu\downarrow}\}$ and
$\sum_\delta
\{
a_{i\downarrow}^\dagger (t) a_{i+\delta\downarrow}(t)
- a_{i-\delta \downarrow}^\dagger (t) a_{i\downarrow}(t)
\} $
by their (vanishing) expectation values in the spin-symmetric,
translationally-invariant ground state.
If $\hat{A}$ represents these operators,
then $\langle \Psi | \hat{A} (t) \hat{A} (t') |\Psi \rangle $
and  $\langle \Psi | \hat{A} (t)^2  |\Psi \rangle $
are in general non-vanishing due to fluctuations and correlations
present in the ground state.
In the following we approximately account for
these fluctuations by replacing the operator
$\hat{A}$ by static, independently random terms
$A_i $ distributed across the lattice with
$\langle A_i \rangle =0$ and
$\langle A_i ^2 \rangle $ determined self-consistently
from $\langle \Psi | \hat{A} (t)^2  |\Psi \rangle $.

When the above replacements are made, we  see from Eq. (20)
that the charge/potential fluctuation terms
$U \langle n_{i\downarrow}+n_{i+\lambda\downarrow} -n \rangle$
and 
$U \langle n_{i\downarrow} - n_{i+\mu\downarrow} \rangle $
couple with the hopping term, leading to charge fluctuation
induced hopping.
On the other hand, the hopping induced fluctuation term
$\sum_\delta
t_\delta \langle a_{i\downarrow}^\dagger (t) a_{i+\delta\downarrow}(t)
- a_{i-\delta \downarrow}^\dagger (t) a_{i\downarrow}(t) \rangle $
is purely imaginary and couples with the local interaction term
yielding a hopping induced relaxation.

\subsection{Charge fluctuation induced hopping}
In Eq. (20)  the operators 
$\{n_{i\downarrow}+n_{i+\lambda\downarrow} -n \}$
and $\{n_{i\downarrow} - n_{i+\mu\downarrow}\}$
were replaced (in the decoupling procedure)
by their expectation values, which vanish in 
the paramagnetic state at half filling ($n=1$),
and also when AF correlations are present.
As shown below this approximation 
essentially amounts to a neglect of
local charge fluctuations
and is valid in the low-temperature, strong-coupling limit,
when double occupancy and vacancy are strongly suppressed.
In the strong coupling limit
we have a spin picture of the ground state,
and for arbitrary orientation of the local spin direction,
the ground state expectation values
$\langle n_{i\downarrow}  + n_{i+\lambda\downarrow} \rangle$
and $\langle n_{i\downarrow} - n_{i+\mu\downarrow} \rangle $
remain invariant
due to the AF correlations between neighbouring spins.
If the spin at site $i$ is inclined at angle $\theta$
from the $-z$ direction, then
$\langle n_{i\downarrow}  + n_{i+\lambda\downarrow} \rangle =
\cos ^2 \theta/2 + \sin ^2 \theta/2 =1 $,
and
$\langle n_{i\downarrow} - n_{i+\mu\downarrow} \rangle = 0$.
Hence the operators
$ n_{i\downarrow} + n_{i+\lambda\downarrow}$
and $ n_{i\downarrow} - n_{i+\mu\downarrow}$
essentially behave like  c-numbers,
and the above replacements are valid.
Even the strong quantum spin fluctuations, which
significantly reduce the AF order parameter in two
dimensions, do not substantially alter this picture
as the major spin-fluctuation contribution arises from
long-wavelength magnon modes.

However, for weak correlation and/or high temperature,
when double occupancy is not strongly penalized,
the local densities
$\langle n_{i\downarrow} \rangle$ and 
$\langle n_{i+\delta\downarrow} \rangle$
{\em independently} fluctuate between zero and one
at short time scales, and therefore
$\langle n_{i\downarrow}+n_{i+\lambda\downarrow} -n \rangle$
and $\langle n_{i\downarrow}-n_{i+\mu\downarrow}\rangle$
strongly fluctuate between minus one and plus one.
In the following we account for these fluctuations
within a static random approximation,
together with disorder averaging at the CPA level.
A more complete theory which includes the spatial and
temporal correlations in the fluctuations will be
developed later by considering the appropriate
two-body Green's functions in Eq. (20) within systematic
approximations like the RPA.

Even in the absence of charge fluctuations,
$\langle n_{i\downarrow}+n_{i+\lambda\downarrow} -n \rangle$
can fluctuate if ferromagnetic correlations are present 
in the (paramagnetic) ground state.
Similarly these fluctuations are also present if the
AF correlations are only short-ranged, so that 
the neighbouring spins are twisted from perfect AF alignment.
This is further discussed in section VIII.

We note that the energy scale $Ut$ of these fluctuation terms
is intermediate between the local $U^2$ scale
and the second-order hopping scale $t^2$ in Eq. (14).
The effect of these  order-$Ut$ fluctuation terms 
is therefore to renormalize the bandwidth by order
$t$ in the strong coupling limit,
as opposed to the order $J$ renormalization by
the second-order virtual hopping process.
Thus the charge-fluctuation and spin-twisting processes
activate the hopping term at first order. 

For simplicity, we consider the NN hopping case, so that
the sites $i$ and $i+\delta$ are in opposite sublattices. 
We assume that $U$ times the density fluctuation term
$U \langle n_{i\downarrow}+n_{i+\delta\downarrow} -n \rangle $,
which we denote by $ V_{i\delta} $, 
is a random potential distributed independently across the lattice.
The average of this random potential term vanishes, but its
second moment $\langle V_{i\delta} ^2 \rangle =U^2
\langle (n_{i\downarrow}+n_{i+\delta\downarrow} -n)^2 \rangle $
reflects a measure of fluctuations in the system.
At half filling ($n=1$), the long-time average yields the
fluctuation magnitude $\delta n ^2 \equiv 
\langle (n_{i\downarrow}+n_{i+\delta\downarrow} -n)^2 \rangle =
2\langle n_{i\downarrow} n_{i+\delta\downarrow} \rangle $,
which has the following limits. 
In the strong coupling limit when double occupancy is strongly
suppressed and AF correlations are present,
we have 
$\langle n_{i\downarrow} n_{i+\delta\downarrow} \rangle
=\langle n_{i\downarrow} n_{i\uparrow} \rangle = 0 $.
At the other extreme, in the non-interacting limit,
when a site may be independently occupied or vacant with
equal probability of one-half, we have 
$2\langle n_{i\downarrow} n_{i+\delta\downarrow} \rangle
=1/2 $. Therefore, the fluctuation magnitude lies in the range 
$0 \le  \delta n ^2 \le 1/2$. 

The random potential term $V_{i\delta}$ is treated at the CPA level,
and the disorder-averaged Green's function is obtained in the 
following.
After the same decoupling approximation made earlier,
we rewrite Eq. (14) in a matrix representation 
(constructed from the site basis) as
\begin{equation}
[\omega^2 {\bf 1} -(\Delta^2 {\bf 1} + {\bf T}^2) 
+ t{\bf V}) ] {\bf G}(\omega)=\omega {\bf 1} + {\bf T}  
\; ,
\end{equation}
where the matrix ${\bf T}$ is associated with the NN hopping term 
and has matrix elements ${\bf T}_{i,i+\delta}=-t$,
whereas ${\bf V}$ 
has matrix elements ${\bf V}_{i,i+\delta}= V_{i\delta}$,
the random potential term.
For further convenience we introduce another Green's function
${\bf g}(\omega)$
through the relation ${\bf G}(\omega)={\bf g}(\omega)
.[\omega {\bf 1} + {\bf T}] $, so that 
${\bf g}(\omega)$ obeys the equation,
\begin{equation}
[\omega^2 {\bf 1} -(\Delta^2 {\bf 1} + {\bf T}^2) 
+t{\bf V}]{\bf g}(\omega)= {\bf 1} \; .
\end{equation}
 
The usual CPA treatment of the random potential term,
involving perturbative expansion followed by disorder averaging,
leads to the following result at the level of rainbow diagrams,
\begin{equation}
\langle {\bf g}(\omega) \rangle=
\frac{ {\bf 1} }
{\omega^2 {\bf 1} -(\Delta^2 {\bf 1} + {\bf T}^2) 
- {\bf \Sigma }(\omega) }
\end{equation}
where,
\begin{equation}
{\bf \Sigma}(\omega)= t^2
\langle {\bf V} {\bf g}(\omega) {\bf V}\rangle
\end{equation}
is the disorder-averaged self energy, to be determined 
self consistently. 

For the purpose of disorder averaging, we assume for 
simplicity that the random potential terms 
$V_{i\delta}$ are independently random
for each pair of sites $i$ and $i+\delta$, so that
$\langle V_{i\delta}. V_{j\delta'}\rangle = \gamma 
(\delta_{ij}\delta_{\delta\delta'} +
\delta_{i,j+\delta'}\delta_{i+\delta,j}) $,
where the second moment $\gamma \equiv 
\langle V_{i,i+\delta} ^2 \rangle $
is a measure of the potential disorder.
This essentially amounts to a neglect of spatial correlations
in the fluctuations. 
In this case there are two terms in the disorder self
energy
\begin{eqnarray}
{\bf \Sigma^1}_{ii}(\omega)  &=&
\gamma t^2 \sum_\delta  
\langle {\bf g} _{i+\delta,i+\delta}(\omega) \rangle \nonumber \\
{\bf \Sigma^2}_{i,i+\delta}(\omega) &=&
 \gamma  t^2 \langle {\bf g} _{i+\delta,i}(\omega) \rangle \; .
\end{eqnarray}
The second term 
${\bf \Sigma^2}$, however,  vanishes because
$\langle {\bf g}_{i+\delta,i}\rangle =\sum_{\bf k'} 
\langle g({\bf k'})\rangle
e^{i{\bf k' . r_\delta} } $ is identically zero.
This follows because $\langle g(\bf k')\rangle $,
given by the Fourier transform of Eq. (25),
is even under the transformation
${\bf k'}\rightarrow {\bf k'}+ \mbox{\boldmath $\pi$} $, 
while the other term 
$e^{i{\bf k' . r_\delta} } $
is odd, and  
therefore the $\bf k'$ summation over the whole Brillouin zone 
yields the vanishing result. 
The disorder self energy is therefore
diagonal in the site basis, and hence
momentum independent. Furthermore, the sum
$\sum_\delta$ in the equation for 
${\bf \Sigma^1}_{ii}(\omega)$ yields $z$, the number of 
nearest neighbours, so that the self energy involves the 
combination $zt^2$. 
These features of the local self energy
are as in the DMFT where the hopping term is taken to be
$t=t^\ast /\sqrt{z}$, so that $zt^2 = (t^\ast)^2 $ is finite
in the limit of infinite dimensions.\cite{vollhardt}

Finally, after Fourier transformation of Eq. (25) 
and of $\langle {\bf G}(\omega)\rangle =
\langle {\bf g}(\omega)\rangle 
.[\omega {\bf 1} + {\bf T}] $, we obtain the 
following result for the full Green's function
\begin{equation}
\langle G({\bf k}\omega) \rangle =
\frac{\omega+ \epsilon_{\bf k}}
{\omega^2 -(\Delta^2 + \epsilon_{\bf k} ^2) -
\Sigma(\omega)} \; ,
\end{equation}
where the disorder-averaged self energy is obtained self consistently from
\begin{equation}
\Sigma(\omega)= z \gamma t^2 \sum_{\bf k'} 
\frac{1}{\omega^2 -(\Delta^2 +\epsilon_{\bf k'} ^2 ) 
-\Sigma(\omega)} \; .
\end{equation}
The above provides a self-consistent scheme for determining the
Green's function when fluctuations are included.
In the following we quantitatively examine the fluctuation
effects on the spectral properties, 
in particular the charge gap and the density of states.
Thermal contribution to charge and spin fluctuations 
brings in temperature into the problem, 
and makes the charge gap sensitive to temperature,
leading to the possibility of a
temperature-driven metal-insulator transition.
Temporal and spatial correlations in fluctuations are neglected in this
treatment.
The present study is nonetheless useful
because when these correlations are included by
considering the appropriate two-body Green's functions
(having the form $\Sigma . G$), the structure of the resulting
equation for $G$ will be similar to Eq. (28).

\subsection{Hopping induced relaxation}
The purely imaginary hopping induced fluctuation term
$U \sum_\delta
t_\delta \langle a_{i\downarrow}^\dagger (t) a_{i+\delta\downarrow}(t)
- a_{i-\delta \downarrow}^\dagger (t) a_{i\downarrow}(t) \rangle
\equiv i V_i '$
couples diagonally to the Green's function.
Therefore, Eq. (23) will contain,
in addition, a diagonal term $t i {\bf V'}$, where
$({\bf V'})_{ij}=\delta _{ij} V_i '$.
Again assuming that $V_i '$ are independently random,
and are uncorrelated with the $V_{i\delta}$,
within the CPA we obtain another diagonal self energy
with a negative coefficient which will simply renormalize
the $\gamma$ defined earlier.

\section{Charge gap and the metal-insulator transition}
A qualitative criterion for the metal-insulator transition
is obtained by considering the competition between the
kinetic energy term (bandwidth $W$) which delocalizes the
electrons making the system metallic,
and the correlation term (Coulomb barrier $U$) 
which localizes the electrons making the system insulating. 
We will first make a simple analysis of the 
effective charge gap when fluctuations are
included to their full extent. We shall see that the
qualitative criterion $U_c  \approx W$ for the vanishing
of the gap, and therefore for the metal-insulator transition,
arises in a very natural way.  

We have seen earlier that when fluctuations are allowed to 
their full extent, 
$ \delta n ^2 = 1/2$,
and therefore the disorder strength $\gamma =U^2 /2$.
Considering the disorder self energy from Eq. (29)
to the lowest order, we have for $\omega =0$
\begin{equation}
\Sigma(0)= \frac{z t^2 U^2}{2} \sum_{\bf k'} 
\frac{1}{ -(\Delta^2 +\epsilon_{\bf k'} ^2 ) }
\sim -2 z t^2 \; ,
\end{equation}
where the result at the right is obtained by 
neglecting the free-electron energy $\epsilon_{k'}$
in comparison to $\Delta = U/2$, for the purpose of obtaining
an estimate. 
Now from Eq. (28) the charge gap vanishes when 
$|\Sigma(0)| = \Delta^2$,
which leads to the following estimate for the 
critical value $U_c$ for the metal-insulator transition,
\begin{equation}
U_c = \sqrt{8zt^2} = 2\sqrt{2} \; t^\ast
\end{equation}
which is of the order of the free-electron band width.
Within the approximations used,
for $U>U_c$ the energy gap never closes,
and the system remains insulating.

We now obtain the self-consistent solution for the self energy
$\Sigma(\omega)$ from Eq. (29). As $\Sigma(\omega)$ is 
complex in general, 
it is convenient to introduce a complex function
\begin{eqnarray}
F(\alpha,\beta) &=&
{\cal R}(\alpha,\beta) - i \beta {\cal I}(\alpha,\beta)
= \sum_{\bf k'} 
\frac{1}{\epsilon_{\bf k'} ^2 + (\alpha + i \beta)} 
\nonumber \\
&=& \sum_{\bf k'} 
\frac{\epsilon_{\bf k'} ^2 + \alpha}
{(\epsilon_{\bf k'} ^2 + \alpha)^2 + \beta^2} 
- i \sum_{\bf k'} 
\frac{\beta}
{(\epsilon_{\bf k'} ^2 + \alpha)^2 + \beta^2}  \;  . 
\end{eqnarray}
By equating $(\Delta^2-\omega^2)
+\Sigma(\omega) = \alpha + i \beta$ in Eq. (29)
the self-consistency equation for $\Sigma(\omega)$ may be recast
in terms of $\alpha,\beta$;
if $\Gamma \equiv z\gamma t^2$ denotes the prefactor,
Eq. (29) then reads
$-\Sigma(\omega)=\Gamma F(\alpha,\beta)$, the real and imaginary
components of which are 
\begin{eqnarray}
\Delta^2 - \omega^2 &=& \Gamma {\cal R}(\alpha,\beta) + \alpha  \\
\beta &=&  \Gamma \beta {\cal I}(\alpha,\beta) \; .
\end{eqnarray}

We now discuss the solution for the above two coupled, non-linear
equations in $\alpha$ and $\beta$, where the parameter
$\Gamma = z\gamma t^2 = z t^2 U^2
\langle (n_{i\downarrow}+n_{i+\delta\downarrow} -n )^2 \rangle =
z t^2 U^2 \delta n ^2 $
incorporates the fluctuation magnitude.
We first consider the solution with $\beta=0$,
which corresponds to a vanishing imaginary part of the self energy
and the density of states, and therefore to the energy gap region.
Since ${\cal I}(\alpha,0)$ decreases monotonically
with positive $\alpha$, for a given $\Gamma$

\begin{figure}
\vspace*{-60mm}
\hspace*{-28mm}
\psfig{file=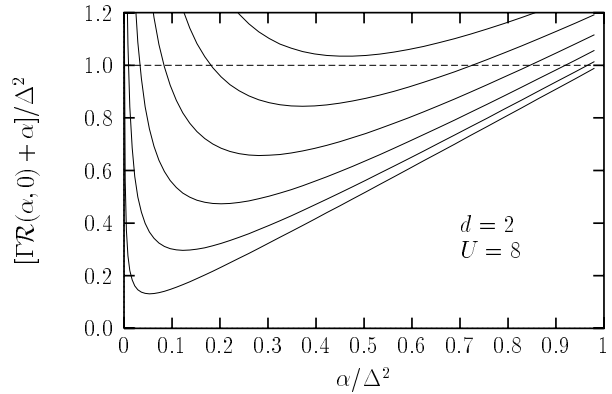,width=135mm,angle=0}
\vspace{-70mm}
\caption{Plot of $\Gamma {\cal R}(\alpha,0) + \alpha $ vs.
$\alpha$, both in units of $\Delta^2$,
for different values of the fluctuation strength $\delta n =$
0.1 (lowest), 0.2, 0.3, 0.4, 0.5, and 0.6 (highest).
The minimum corresponds to a $\omega$ value from Eq. (33)
which yields the band edge. As the minimum approaches 1
from below, the band gap shrinks to zero.}
\end{figure}

\noindent
there exists
a physically significant 
solution $\alpha=\alpha^\ast > 0$ of the equation
$1=\Gamma {\cal I}(\alpha,0)$.
For $\alpha > \alpha^\ast$, we have 
${\cal I}(\alpha,\beta) < 
{\cal I}(\alpha,0) < {\cal I}(\alpha^\ast,0)$,
therefore Eq. (34) has a solution only if $\beta=0$,
whereas for $\alpha < \alpha^\ast$ a solution exists
only for $\beta \ne 0$. 
Thus, $\alpha^\ast$ is the point where the self energy
changes from purely real to complex, and therefore corresponds
to the band-edge energy, as further discussed below.

There is another special value $\alpha=\alpha^\dagger$,
this one having to do with Eq. (33). 
The right-hand-side term 
$\Gamma {\cal R}(\alpha,0) + \alpha$
diverges in both the limits
$\alpha \rightarrow 0$  and $\alpha \rightarrow \infty$, 
and therefore it must have a minimum at some intermediate value
$\alpha=\alpha^\dagger$,
as shown in Fig. 1 from an explicit calculation. 
Since $\Gamma {\cal R}(\alpha,0) + \alpha 
=\Delta^2 - \omega^2$,
this minimum corresponds to a maximum value 
$\omega=\omega^\dagger$ beyond which no real solution for 
$\omega$ exists. 

It is interesting to note that these two values 
$\alpha^\ast$ and $\alpha^\dagger$ are actually {\em identical}
because their parent equations $\Gamma dR/d\alpha + 1 = 0$ 
and $1-\Gamma {\cal I}=0$ 
are so in view of the relation $dR/d\alpha = -{\cal I}$.
The significance of this coincidence is that the imaginary
part of the self energy grows continuously,
so that there is no discontinuity in the density of states, 
and therefore when the band gap vanishes 
the resulting metal-insulator transition is also continuous.

Solving for real $\omega$ from Eq. (33) for a given 
$\alpha > \alpha^\ast$ and $\beta=0$, 
we thus have the solution to the Eqs. (33), (34).
If $\omega^\ast$ is the solution corresponding to
$\alpha=\alpha^\ast$, then 
since $\alpha^\ast = \alpha^\dagger$,
we have $\omega^\ast = \omega^\dagger$,
beyond which no real solution for $\omega$ exists,
as discussed earlier. 
Therefore $\omega^\ast$ is an upperbound for 
solutions with $\beta=0$,
for which the imaginary part of the self energy and
the density of states vanish.
This implies that $\omega^\ast$ is the band-edge energy, 
and $\omega< \omega^\ast$  is the energy gap region, 
with the energy gap
$E_g(\Gamma)=2\omega^\ast(\Gamma)$.

\begin{figure}
\vspace*{-60mm}
\hspace*{-28mm}
\psfig{file=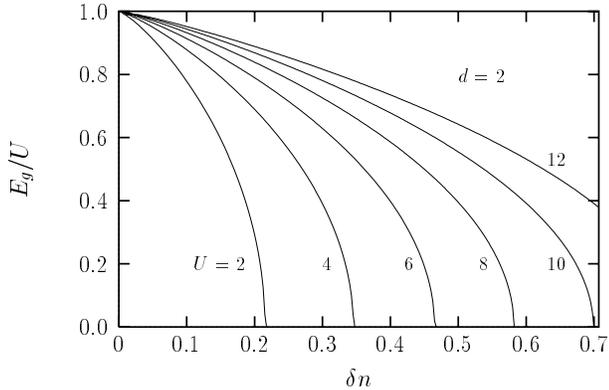,width=135mm,angle=0}
\vspace{-70mm}
\caption{The normalized energy gap $E_g/U$ in two dimensions
vs. the fluctuation strength in the range
$0 \le \delta n ^2 \le 1/2 $ for different $U$ values.
The critical $U$ value, above which
the band gap never closes, is slightly above 10.}
\end{figure}

A convenient way to obtain $\omega^\ast$,
and hence the band gap
$E_g(\Gamma)$ is from the plot of 
$\Gamma {\cal R}(\alpha,0) + \alpha$ vs. $\alpha$ 
in the range $0 <\alpha < \Delta^2  $.
The variation of the band gap so obtained with 
the fluctuation magnitude $\delta n $
in the range $0<\delta n^2 < 1/2$ 
is shown for two and three dimensions
in Figs. 2 and 3 respectively for different values of $U$.
There exists a critical interaction strength $U_c$, 
such that for $U<U_c$ with increasing  
$\delta n $  the energy gap vanishes at some critical value 
$\delta n_c (U) $, whereas for $U>U_c$
the band gap remains finite and the system remains insulating.
In two and three dimensions 
we obtain the critical interaction strengths
$U_c = 10.2$ and $12.5$ respectively,
($U_c/W = 1.27$  and $1.04$). 

With increasing fluctuation strength $\Gamma$ 
the solution $\alpha^\ast$ 
of the equation $1=\Gamma {\cal I}(\alpha,0)$
increases, as also the right-hand side of Eq. (33),
so that the corresponding $\omega^\ast$
decreases until at some critical value $\Gamma=\Gamma^\ast$
we finally have $\omega^\ast =0$, and the left-hand
side of Eq. (33) cannot increase any further.
This critical fluctuation strength at which 
$\omega^\ast$ and hence the band gap
vanish marks the metal-insulator transition.
For $\Gamma>\Gamma^\ast$ a solution of Eq. (33) exists
only if $\beta \ne 0$, which leads to a finite
density of states at $\omega=0$.
It is important to note that $\beta$ and 
the density of states increases
continuously with  $\Gamma-\Gamma^\ast$,
signifying a continuous metal-insulator transition. 

While the range $\alpha>\alpha^\ast$ corresponds to
the energy-gap region with $\beta=0$, 
the other side $\alpha < \alpha^\ast$ necessarily goes 
with $\beta \ne 0$, and therefore corresponds to the band 
region wherein the self energy is complex. 
It is also clear from Eq. (33) that for $\omega^2 > \Delta^2$
a solution exists only if $\alpha$ is negative, so that
the band region corresponds to $\alpha < 0$.
To solve Eqs. (33), (34) for $\alpha,\beta$ for some fixed $\Gamma$, 
we first solve for $\beta$ from Eq. (34) for a given
$\alpha$, and then solve for a real $\omega$ from Eq. (33)
with this pair $\alpha,\beta$. 

\begin{figure}
\vspace*{-60mm}
\hspace*{-28mm}
\psfig{file=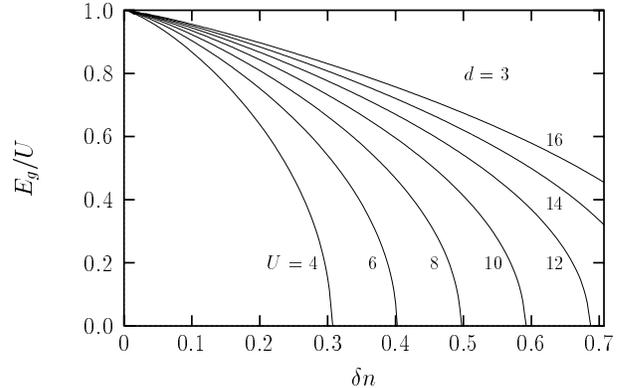,width=135mm,angle=0}
\vspace{-70mm}
\caption{The normalized energy gap $E_g/U$ in three dimensions
vs. the fluctuation strength in the range 
$0 \le \delta n ^2 \le 1/2$
for different $U$ values. The critical $U$ value, above which
the band gap never closes, is slightly above 12.}
\end{figure}

\noindent
By varying $\alpha$ between a sufficiently negative value and 
$ \alpha^\ast$, the full range of solutions are thus obtained
for $\omega>\omega^\ast$. 
From these self-consistent solutions for $\alpha,\beta$ 
the complex self energy is then obtained from 
$(\Delta^2 -\omega^2)+\Sigma(\omega) = \alpha + i \beta$.

\subsection{Density of states}
We now discuss the evaluation of the density of states
using this complex self energy $\Sigma(\omega)$.
For this purpose we rewrite Eq. (28) for the Green's function
as 
\begin{equation}
G({\bf k},\omega)= \frac{\omega+\epsilon_{\bf k}}
{\omega^2 - {\cal E}^2 ({\bf k},\omega) } = \frac{1}{2}
\left [ \frac{1+\frac{\epsilon_{\bf k}}{\cal E}}
{\omega-{\cal E}} + 
\frac{1-\frac{\epsilon_{\bf k}}{\cal E}}
{\omega+{\cal E}} \right ]
\end{equation}
where ${\cal E}^2 ({\bf k},\omega) =
E_{\bf k}^2 + \Sigma(\omega)$.
Writing 
${\cal E}({\bf k},\omega) =
|{\cal E}({\bf k},\omega)| e^{i\theta ({\bf k},\omega) }$,
from the solutions
for $|{\cal E}({\bf k},\omega)|$
and $\theta ({\bf k},\omega)$
\begin{eqnarray}
|{\cal E}({\bf k},\omega)| &=& 
[\{E_{\bf k}^2 + \Sigma_r(\omega) \}^2 + \Sigma_i^2(\omega) 
]^{1/4} \nonumber \\
\theta({\bf k},\omega) &=& \frac{1}{2}\tan ^{-1} 
\frac{\Sigma_i(\omega)}
{E_{\bf k}^2 + \Sigma_r(\omega)} 
\end{eqnarray}
we obtain ${\cal E}({\bf k},\omega)=
|{\cal E}({\bf k},\omega)|e^{i\theta({\bf k},\omega)}
={\cal E}_r ({\bf k},\omega) + i {\cal E}_i ({\bf k},\omega)$, 
in terms of which the density of states is finally obtained from
\begin{equation}
N(\omega)=\frac{1}{2\pi}
\sum_{\bf k} \left [ \frac{{\cal E}_i}
{(\omega-{\cal E}_r )^2 + 
{\cal E}_i ^2 } +
\frac{{\cal E}_i}
{(\omega + {\cal E}_r )^2 + 
{\cal E}_i ^2 } \right ] \; .
\end{equation}
Here we have made use of the transformation property 
of $\epsilon_{\bf k}$ which changes sign under the transformation
${\bf k}\rightarrow {\bf k} + \mbox{\boldmath $\pi$}$
while ${\cal E}({\bf k},\omega)$ does not, so that when summed
over all ${\bf k}$ the 
$\epsilon_{\bf k}$ term in Eq. (35) 
yields a vanishing contribution.

\begin{figure}
\vspace*{-60mm}
\hspace*{-28mm}
\psfig{file=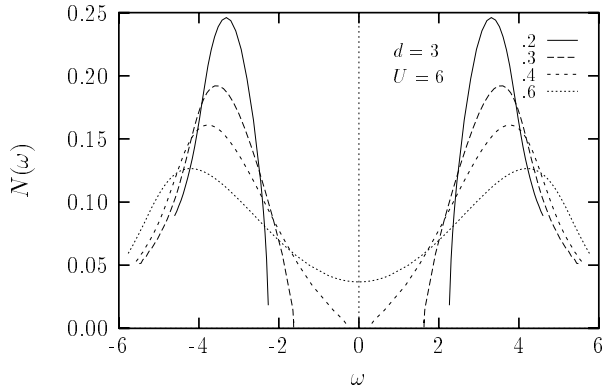,width=135mm,angle=0}
\vspace{-70mm}
\caption{The electronic density of states $N(\omega)$
for different values of the fluctuation strength $\delta n$.}
\end{figure}

The results  for the density of states
for a simple cubic lattice and $U=W/2=6$
are shown in Fig. 4
for several values of the fluctuation strength
$\delta n $. 
Due to the fluctuation-induced states in the gap, 
the band gap decreases with 
$\delta n $, and at a critical value 
$\delta n_c \approx 0.4 $, 
the band gap vanishes and the system
undergoes a continuous metal-insulator transition.

\section{Broken-symmetry state}
So far we have been concerned with the spin-symmetric state. 
However, the present approach provides for 
a description of the broken-symmetry state as well,
and in this section we briefly make contact
with the known results for the AF state
with respect to the AF band energies and amplitudes.  

We consider the NN hopping case, and proceed with the
decoupling approach A. In the AF state the density
$\langle n_{i\downarrow} \rangle$ has values $(n-m)/2$
and $(n+m)/2$ on the two sublattice sites, where $m$ is the
sublattice magnetization. Translational symmetry 
therefore exists only within the sublattice basis.
After employing the same decoupling procedure
as before leading to Eq. (15), Fourier transformation 
within the sublattice basis leads to
\begin{equation}
G({\bf k}\omega)=\frac{1}{\omega^2 - E_{\bf k}^2 }
\left [ \begin{array}{lr}
\omega-\frac{mU}{2} & \epsilon_{\bf k} \\
\epsilon_{\bf k}    & \omega+\frac{mU}{2} 
\end{array} \right ] \; .
\end{equation}
The quasiparticle band energies in the AF state are thus
given by $\omega = \pm E_{\bf k}$, same as for the symmetric case.
To make contact with the HF results for the AF state, we make the 
following approximation consistent with the HF scheme
\begin{equation}
\left \langle \left ( n_{i\downarrow}-\frac{n}{2}\right )^2 
\right \rangle
\approx 
\left \langle  n_{i\downarrow}-\frac{n}{2} 
\right \rangle ^2 = \frac{m^2}{4} \; .
\end{equation}
In this case $\Delta^2 \equiv U^2
\langle ( n_{i\downarrow}-n/2  )^2 \rangle = U^2 m^2 /4$,
so that $2\Delta = mU$, which now involves the sublattice magnetization. 
With this approximation, Eq. (38) simplifies to 
\begin{equation}
G({\bf k}\omega)=\frac{1}{\omega^2 - E_{\bf k}^2 }
\left [ \begin{array}{lr}
\omega-\Delta & \epsilon_{\bf k} \\
\epsilon_{\bf k}    & \omega+\Delta 
\end{array} \right ]  \; ,
\end{equation}
with $E_{\bf k}=\sqrt{\Delta^2 + \epsilon_{\bf k} ^2 }$.
This is precisely of the same form as obtained 
in the HF description of the AF state.\cite{singh.tesanovic}
Consequently, the
expressions for the AF-state electron amplitudes are identical,
resulting in the same self-consistency condition
$1/U=\sum_{\bf k} 1/2E_{\bf k} $, which determines the 
sublattice magnetization $m$.  

\section{Spin correlations and the electronic spectrum}
AF correlations have already been considered while introducing
the decoupling schemes A, B, and C in section IV.
Here we examine the influence of short-range
AF ordering on such aspects of the electronic spectrum 
as the charge gap, the density of states etc. 
In the strong correlation limit and for half-filling ($n=1$),
when charge fluctuations are absent, the expectation value
$\langle n_{i\downarrow} + n_{i+\delta \downarrow} -n \rangle$
vanishes only for perfect AF ordering when 
$\langle n_{i+\delta\downarrow} \rangle=
\langle n_{i\uparrow} \rangle$.
However, for $T>T_{\rm N}$,
when the spin ordering is not perfect,
the finite spin correlation length $\xi$
results in a twisting of neighbouring spins
by $\pi/\xi$ on the average.
If at site $i+\delta$ the spin is pointing up, 
but at site $i$ the spin instead of pointing
down is twisted by a small angle $\theta$,  then 
$\langle n_{i\downarrow} \rangle =\cos^2 (\theta/2)\approx
1-\theta^2 /4$, so that 
$\langle n_{i\downarrow} + n_{i+\delta \downarrow}\rangle$
is slightly reduced from 1. Whereas if the spin at site $i$ is
pointing down, but at the neighboring site $i+\delta$
the spin instead of pointing up is twisted by the small angle 
$\theta$, then    
$\langle n_{i\downarrow} + n_{i+\delta \downarrow}\rangle$
is slightly increased from 1.
This results in a twist-induced fluctuation term arising from 
the expectation value 
$\langle n_{i\downarrow} + n_{i+\delta \downarrow} -n \rangle$
which fluctuates between $-\theta^2/4$ and $\theta^2/4$, so that
\begin{equation}
|\delta n_\xi | \equiv 
|\langle n_{i\downarrow} + n_{i+\delta \downarrow} -n \rangle |
\approx \frac{\theta^2}{4} = \frac{\pi^2}{4\xi^2} \; .
\end{equation}

This twist-induced fluctuation term couples with the
hopping term at first order in Eq. (14), 
so that the quasiparticle band energies are modified 
by an order-$t$ term. 
This feature is similar to that in the spiral state 
of the doped Hubbard model,
where the twisting of neighbouring spins
activates the hopping term at first order
and broadens the bands symmetrically.\cite{spiral1,spiral2,spiral3}
The pulled-down states are filled whereas the
pushed-up states are empty due to doping,
resulting in stabilization of the spiral state. 

This twist-induced fluctuation term will also contribute to
the overall fluctuation strength discussed in section V,
with the disorder strength $\Gamma$
now including a term related to the
inverse spin-correlation length.   
Therefore with increasing temperature the enhanced magnon
excitation and the spin-fluctuation
induced spin disordering
will also contribute to states in the gap and 
the band-gap reduction. Thus in general, both charge
and spin fluctuations contribute towards an  
effective disorder strength $\Gamma$
which will determine the overall electronic spectral properties.

\section{Doping and transfer of spectral weight}
In the strong coupling limit, doping with holes or electrons
results in transfer of spectral weight between the two Hubbard bands.
Consider the atomic limit ($t=0$) where this feature
is most pronounced and easily understood.
A site with a single electron of spin $\sigma$ contributes two states
at energies $\epsilon_\sigma = 0$ (occupied) and 
$\epsilon_{\overline{\sigma}} = U$ (unoccupied).
When a spin-$\sigma$ hole is introduced by removing the
spin-$\sigma$ electron, the spectrum changes to both states at
energy $\epsilon_\sigma = \epsilon_{\overline{\sigma}} = 0$,
because electrons of either spin can now go into the empty site
with zero energy. Extending to a collection of sites,
if $x$ is the hole concentration, then the density of states
(both spins) at energies $0$ and $U$ are $1+x$ and $1-x$
respectively, so that the density of unoccupied states
in the lower energy level is $2x$.
Finite hopping introduces a spread in the hole energy of order
$t$, so that this picture of the transfer of spectral weight
remains valid as long as the band separation $(\sim U)$ is
much bigger than the bandwidth $(\sim t)$.

In order to examine the effect of hole doping on the spectral
weights, we consider Eq. (14) appropriate for the NN hopping case,
and focus on the correlation function term 
$-iU^2 \; \langle \Psi |{\rm T} 
\{n_{i\downarrow}(t)-n/2 \} ^2 a_{i\uparrow}(t)
a_{j\uparrow}^\dagger (0) |\Psi\rangle $,
which reduces to the single-particle Green's function
in the half-filled case $n=1$.
For a total particle density $n=\langle n_{i\uparrow}\rangle +
\langle n_{i\downarrow}\rangle =1-x < 1$,
corresponding to a hole concentration $x$,
the term $(n_{i\downarrow} - n/2)^2$ reduces to
$(n/2)^2 + x\, n_{i\downarrow} $,
which now contains an operator term for finite doping.
However, the doping-induced term 
$-x\, iU^2 \; \langle \Psi |{\rm T} 
n_{i\downarrow}(t) a_{i\uparrow}(t)
a_{j\uparrow}^\dagger (0) |\Psi\rangle $
can be written in terms of $G(t)$ from Eq. (4) as
\begin{eqnarray}
- &x& i U^2 \; \langle \Psi |{\rm T} 
n_{i\downarrow}(t) a_{i\uparrow}(t)
a_{j\uparrow}^\dagger (0) |\Psi\rangle 
= \nonumber \\
&x& U [ i\partial_t G_{ij}(t) - \delta(t) \delta_{ij}
+ \sum_\delta t_\delta G_{i+\delta,j} (t) ] \; ,
\end{eqnarray}
so that the correlation function term
under consideration can be exactly
obtained in terms of the single-particle Green's function $G$.
Decoupling the equation for $G$ as before in section IV A,
we obtain after Fourier transformation,
\begin{eqnarray}
& & [(\omega - nU/2)^2 - \{ (nU/2)^2 + \epsilon_{\bf k} ^2 \}
- x U (\omega-\epsilon_{\bf k} ) ] G({\bf k}\omega) \nonumber \\
&=&
\omega - nU/2 + \epsilon_{\bf k}  - x U \; .
\end{eqnarray}
After simplification and the Hartree shift
$\omega - U/2 \rightarrow \omega$, we obtain:
\begin{equation}
G({\bf k}\omega)= \frac{\omega - xU/2 + \epsilon_{\bf k} }
{\omega^2 - E_{\bf k} ^2 + x U \epsilon_{\bf k} }
\end{equation}
where as before $E_{\bf k} ^2 = (U/2)^2 + \epsilon_{\bf k} ^2$.
The quasiparticle band energies are therefore given by
$\pm {\cal E}_{\bf k} = \pm \sqrt {E_{\bf k} ^2 - xU\epsilon_{\bf k}}$,
which in the strong coupling limit reduce to
$\pm [U/2 - x\epsilon_{\bf k} ] $,
when only terms upto first order in $t/U$ are retained. 
This indicates that doping leads to a qualitative change
in the nature of the band itself,
with the dispersion changing from $\epsilon_{\bf k} ^2 /U$
of order $J$ to $x \epsilon_{\bf k}$,  
resulting in an effective
doping-induced hopping strength and bandwidth of order $xt$.
This is again very different from the Hubbard I ``rigid-band''
result wherein there is only a slight doping-induced modification
of the bandwidth, but no qualitative change in the nature of the band.

We examine the loss of integrated spectral weight in the upper band
due to doping. Integrating the spectral function $A(\omega)$ over the
upper band, and summing over both spins, we obtain 
\begin{eqnarray}
2 \int^{\oplus} d\omega A(\omega) &=&
2 \int^{\oplus} d\omega \frac{1}{\pi} \sum_{\bf k}
{\rm Im} G({\bf k}\omega) \nonumber \\
&=& 
\sum_{\bf k} \frac{{\cal E}_{\bf k} -xU/2 + \epsilon_{\bf k} }
{{\cal E}_{\bf k}} \; .
\end{eqnarray}
Evaluating the momentum sum in the strong coupling limit
by retaining terms only upto order $t^2/U^2$,
we obtain,
\begin{equation}
2 \int^{\oplus} d\omega A(\omega) =
1-x\left ( 1-24\frac{t^2}{U^2}(1-x^2)\right ) \; .
\end{equation}
Upto order $t$ the integrated spectral weight
in the upper band is therefore simply reduced to $1-x$ by doping,
as in the atomic limit. Similarly the weight in the lower band is
enhanced to $1+x$. 

\section{Conclusions}
In conclusion, the new decoupling scheme 
is appropriate for both the spin-symmetric
(paramagnetic metallic and insulating) phases
as well as the broken-symmetry AF phase.
Although restricted to bipartite lattices,
the decoupling approximations
involve only weakly fluctuating operators, resulting in
several improvements over the Hubbard I scheme.  
Independent of magnetic ordering, the scheme yields,
at the lowest order,
the correct strong-coupling bandwidth of order $J$,
a non-zero critical interaction strength (above which the band gap opens)
only if same-sublattice hopping (e.g., NNN hopping) is also present,
and the correct quasiparticle spectral weights (of 1/2 per spin)
in each band for the half-filled model.

Charge and spin fluctuations,
which are manifested in double occupancy and vacancy of sites 
and spin twisting due to short-range AF ordering,
were studied within a static, random approximation.
The self-consistent, CPA-level evaluation of the disorder-averaged
self energy showed a fluctuation-induced band-broadening
and consequent decrease in the band gap.
The critical interaction strength,
above which the band gap remains finite even when the
fluctuations are included to their full extent,
was obtained as 
$U_c/W = 1.27$  and $1.04$ 
in two and three dimensions respectively. 

The solution of the non-linear self-consistent equation
for the disorder self energy shows that with increasing
fluctuation strength the band gap shrinks to zero continuously
and the density of states $N(0)$ between the bands also grows
continuously. 
The monotonic dependence of fluctuation strength on temperature
translates into a continuous metal-insulator transition.
The fluctuation magnitude as well as its
dynamics can be determined from the appropriate correlation
function, and this is presently under investigation.
The dynamics is relevant in view of the switching of the
spectral function between the two Hubbard bands,
and the possibility of a motionally narrowed central peak in
the density of states when the Hubbard switching rate
becomes comparable with the energy separation between the bands.
Quantum Monte Carlo simulations do show a central peak
in the vicinity of the metal-insulator transition.\cite{schlipf}

With doping a quantitatively correct transfer of spectral weight
between the two Hubbard bands in the strong-coupling limit was obtained
at the lowest-order level in the decoupling scheme.
Furthermore, in contrast to the Hubbard I approximation,
a qualitative change in the nature of the band itself was
obtained, with an effective
doping-induced hopping strength and bandwidth of order $xt$.

Extension of the equation of motion approach
to the transverse spin propagator to obtain the
low-energy spin-fluctuation spectrum will provide a
magnetic description of the spin-symmetric state,
and enable one to approach the magnetic phase boundary from
the paramagnetic side.
This, as well as correlations and dynamics of fluctuations,
and their impact on the electronic and magnetic spectrum,
are presently under investigation.

\section*{Acknowledgments}
The author is grateful to Dieter Vollhardt for helpful correspondence.
This work was supported in part by Research Grant (No. SP/S2/M-25/95)
from the Department of Science and Technology, India.

\end{multicols}
\end{document}